\documentclass[aj_pt4]{aastex}

\slugcomment{Accepted manuscript}

\shorttitle{Constraints on a second planet in the WASP-3 system}
\shortauthors{Maciejewski et al.}

\begin{document}

\title{Constraints on a second planet in the WASP-3 system\footnote{Partly based on (1) observations made at the Centro Astron\'omico Hispano Alem\'an (CAHA), operated jointly by the Max-Planck Institut f\"ur Astronomie and the Instituto de Astrof\'{\i}sica de Andaluc\'{\i}a (CSIC), (2) data collected with telescopes at the Rozhen National Astronomical Observatory, and (3) observations obtained with telescopes of the University Observatory Jena, which is operated by the Astrophysical Institute of the Friedrich-Schiller-University.}}

\author{G.~Maciejewski\altaffilmark{1}, 
		A.~Niedzielski\altaffilmark{1}, 
		A.~Wolszczan\altaffilmark{2,3}, 
		G.~Nowak\altaffilmark{1}, 
		R.~Neuh\"auser\altaffilmark{4},
		J.~N.~Winn\altaffilmark{5},
		B.~Deka\altaffilmark{1},
		M.~Adam\'ow\altaffilmark{1},
		M.~G\'orecka\altaffilmark{1},
		M.~Fern\'andez\altaffilmark{6},
		F.~J.~Aceituno\altaffilmark{6},
		J.~Ohlert\altaffilmark{7,8},
		R.~Errmann\altaffilmark{4,9},
		M.~Seeliger\altaffilmark{4},
		D.~Dimitrov\altaffilmark{10},
		D.~W.~Latham\altaffilmark{11},
		G.~A.~Esquerdo\altaffilmark{11},
		L.~McKnight\altaffilmark{5},
		M.~J.~Holman\altaffilmark{11},
		E.~L.~N.~Jensen\altaffilmark{12},
		U.~Kramm\altaffilmark{13},
		T.~Pribulla\altaffilmark{14},
		St.~Raetz\altaffilmark{4},
		T.~O.~B.~Schmidt\altaffilmark{4},
		Ch.~Ginski\altaffilmark{4},
		S.~Mottola\altaffilmark{15},
		S.~Hellmich\altaffilmark{15},
		Ch.~Adam\altaffilmark{4},
		H.~Gilbert\altaffilmark{4},
		M.~Mugrauer\altaffilmark{4},
		G.~Saral\altaffilmark{16,11},
		V.~Popov\altaffilmark{10},
		M.~Raetz\altaffilmark{17}
}

\altaffiltext{1}{Centre for Astronomy, Faculty of Physics, Astronomy and Informatics, 
             Nicolaus Copernicus University, Grudziadzka 5, 87-100 Torun, Poland}
\altaffiltext{2}{Department of Astronomy and Astrophysics, The Pennsylvania State University, 525 Davey Laboratory, University Park, PA 16802, USA}
\altaffiltext{3}{Center for Exoplanets and Habitable Worlds, The Pennsylvania State University, 525 Davey Laboratory, University Park, PA 16802, USA}
\altaffiltext{4}{Astrophysikalisches Institut und Universit\"ats-Sternwarte,  Schillerg\"asschen 2--3, D--07745 Jena, Germany}
\altaffiltext{5}{Department of Physics, Massachusetts Institute of Technology, Cambridge, MA 02139, USA}
\altaffiltext{6}{Instituto de Astrof\'isica de Andaluc\'ia (IAA-CSIC), Glorieta de la Astronom\'ia 3, 18008 Granada, Spain}
\altaffiltext{7}{Michael Adrian Observatorium, Astronomie Stiftung Trebur, 65468 Trebur, Germany}
\altaffiltext{8}{University of Applied Sciences, Technische Hochschule Mittelhessen, 61169 Friedberg, Germany}
\altaffiltext{9}{Abbe Center of Photonics, Max-Wien-Platz 1, 07743 Jena, Germany}
\altaffiltext{10}{Institute of Astronomy, Bulgarian Academy of Sciences, 72 Tsarigradsko Chausse Blvd., 1784 Sofia, Bulgaria}
\altaffiltext{11}{Harvard-Smithsonian Center for Astrophysics, 60 Garden Street, Cambridge, MA 02138, USA}
\altaffiltext{12}{Department of Physics and Astronomy, Swarthmore College, Swarthmore, PA 19081, USA}
\altaffiltext{13}{Institut f\"ur Physik, University Rostock, Rostock, Germany}
\altaffiltext{14}{Astronomical Institute, Slovak Academy of Sciences, 059 60 Tatransk\'a Lomnica, Slovakia}
\altaffiltext{15}{German Aerospace Center (DLR), Rutherfordstra{\ss}e 2, 12489 Berlin, Germany}
\altaffiltext{16}{Astronomy and Space Sciences Department, Istanbul University, 34116 Fatih Istanbul, Turkey}
\altaffiltext{17}{Private observatory, Stiller Berg 6, 98587 Herges-Hallenberg, Germany}

\begin{abstract}

  There have been previous hints that the transiting planet WASP-3~b is accompanied by a second planet in a nearby orbit, based on small deviations from strict periodicity of the observed transits.  Here we present 17 precise radial velocity measurements and 32 transit light curves that were acquired between 2009 and 2011. These data were used to refine the parameters of the host star and transiting planet. This has resulted in reduced uncertainties for the radii and masses of the star and planet.  The radial-velocity data and the transit times show no evidence for an additional planet in the system.  Therefore, we have determined the upper limit on the mass of any hypothetical second planet, as a function of its orbital period.

\end{abstract}



\section{Introduction}

The extrasolar planetary system WASP-3 comprises a transiting planet with a mass of $1.7$ times the mass of Jupiter ($M_{\rm{Jup}}$), which orbits a main-sequence star of spectral type F7-8 and apparent magnitude $V=10.6$ \citep{Pollacco08}. The orbital period is only $1.85$~d, making WASP-3~b one of the hottest planets known at the time of its discovery. The planetary radius, which is known to be 1.3 times larger than the radius of Jupiter ($R_{\rm{Jup}}$), is consistent with atmospheric models for strongly-irradiated giant planets.

The system properties have been determined through several photometric and spectroscopic follow-up studies. \citet{Gibson08}, \citet{Tripathi10}, and \citet{Christiansen11} acquired high-precision transit light curves and redetermined the stellar and planetary radii, orbital inclination, and transit ephemeris. The radial velocities (RVs) measured during transits exhibit the Rossiter--McLaughlin (RM) effect \citep{Rossiter,McLaughlin}, in a pattern that indicates a low value for the sky-projected angle between the stellar spin axis and the planetary orbital axis \citep{Simpson10,Tripathi10,Miller10}.

\citet{Tripathi10} found that the measured mid-transit times seemed to be statistically inconsistent with strict periodicity, i.e., fitting the measured times to a linear function of epoch gave an unacceptably large $\chi^2$. They noted that this could be explained as a consequence of either variations in the planetary orbit due to an unseen companion, or underestimated uncertainties in the measured mid-transit times. \citet{Maciejewski10} measured six mid-transit times from light curves acquired with 0.6-m class telescopes, combined them with data from the literature, and concluded that the transit times may be modulated with a period of $\sim$127 days and a peak-to-peak amplitude of $\sim$4 minutes. Numerical simulations showed that such a pattern of transit timing variation (TTV) could be produced by an additional low-mass planet. The postulated TTV signal has not been confirmed by further observations \citep{Littlefield11,Sada12,Nascimbeni13,Montalto12}, and analyses of the mid-transit times now exclude the originally-postulated periodic TTV signal \citep{Nascimbeni13,Montalto12}. However, \citet{Nascimbeni13} showed that the measured mid-transit times are still not statistically consistent with a linear ephemeris, and pointed out that such apparently chaotic timing variations could be produced by some specific orbital configurations. \citet{Montalto12} suggested that chromospheric activity of the parent star could be a potential source of the transit timing noise.

In this paper we present results of new photometric and spectroscopic follow-up observations, the goal of which was to confirm or refute the hypothesis about the second planet in the WASP-3 system. As our newly measured mid-transit times are consistent with a linear ephemeris, and no signal in RV residuals from a single planet orbital fit is detected, we determine upper limits on the mass of a hypothetical second planet as a function of its orbital period. In addition, we redetermined the stellar and planetary properties of the WASP-3 system using our spectroscopic and photometric data sets.


\section{Observations and data reduction}

\subsection{Transit photometry}\label{Sect.TransitObs}

Twenty-six transits of WASP-3~b were observed between 2009 September and 2011 October. Four of these transits were observed with two different telescopes, and one of the transits was observed with three different telescopes, giving a total of 32 light curves.  Most of these data were collected with telescopes with diameters greater than 1-m, enabling a photometric precision better than 1.5~mmag. A portion of the data was obtained in collaboration with the Young Exoplanet Transit Initiative \citep[YETI,][]{Neuhauser11}. Individual observations are summarised in Table~\ref{Tab.LCObservations}, and the light curves are plotted in Fig.~\ref{Fig.LCs}.\footnote{The photometric data are available at {\tt http://ttv.astri.umk.pl}} Short descriptions of the instrumental set-up and the observations are given below, arranged in descending order of telescope aperture size.

Four complete transits were observed in 2010 with the 2.2-m telescope at the Calar Alto Observatory (CAO, Spain) within the programme H10-2.2-010.  For the detector, we used the Calar Alto Faint Object Spectrograph (CAFOS) in its imaging mode. The full field of view (FoV) was windowed to record the target star and nearby bright comparison star. Binning of $2\times2$ was applied to shorten the read-out time. The optical set-up was significantly defocused to reduce the impact of flat-fielding imperfections. The telescope was autoguided to hold stellar images fixed on the same pixels throughout each night. The weather conditions on 2010 Aug 29 were non-photometric due to thin clouds, leading to reduced quality of the time-series photometry. During the other nights the sky was mainly clear but not perfect enough to achieve sub-millimagnitude precision. The gaps in light curves on 2010 Aug 05 and Sep 11 were caused by passing clouds.

The 2.0-m Ritchey-Chr\'etien-Coud\'e telescope at the Bulgarian National Astronomical Observatory Rozhen was used to observe a transit on 2010 Jun 07.  The detector was a Roper Scientific VersArray 1300B CCD camera ($1340 \times 1300$ pixels, FoV: $5\farcm8 \times 5\farcm6$).

Observations of five transits were carried out in 2011 with the 1.5-m telescope at the Sierra Nevada Observatory (OSN) operated by the Instituto de Astrof\'isica de Andaluc\'ia (Spain). A Roper Scientific VersArray 2048B CCD camera ($2048 \times 2048$ pixels, FoV: $7\farcm92 \times 7\farcm92$) was used as a detector. The observations on 2011 Jun 15 were interrupted by technical failures. The lower quality data from 2011 Aug 26, especially at the end of the run, were a consequence of observing through a large airmass.

The 1.2-m Trebur Telescope at the Michael Adrian Observatory in Trebur (Germany) was used to observe a transit in 2010 and four events in 2011. The Cassegrain-type telescope is equipped with a $3072\times2048$-pixel SBIG STL-6303 CCD camera (FoV: $10\farcm0 \times 6\farcm7$). A $2\times2$ binning mode was used. 

Four transits were observed with the 1.2-m telescope at the Fred Lawrence Whipple Observatory (FLWO, USA). The detector was the Keplercam CCD system ($4096\times 4096$~pixels, binned $2\times 2$, FoV: $23\farcm \times 23\farcm$).  The observations on 2010 Jun 26 were interrupted by occasionally passing clouds.

Observations of a transit on 2011 Aug 26 were conducted with the 1.23-m telescope at CAO. A CCD camera equipped with an e2v CCD231-84-NIMO-BI-DD sensor ($4096\times4096$ pixels) was used, providing a $12\farcm5 \times 11\farcm5$ FoV.

The 0.9/0.6-m Schmidt telescope at the University Observatory Jena in Gro{\ss}schwabhausen near Jena (Germany), equipped with the CCD-imager Schmidt Teleskop Kamera \citep[STK,][FoV: $52\farcm8 \times 52\farcm8$]{mugrauer}, was used to observe three transits. All light curves were acquired with clear sky, but some portion of data was affected by adverse weather conditions (haze and high humidity).

Four light curves were taken at the Peter van de Kamp Observatory at Swarthmore College (Swarthmore, PA, USA) with an 0.6-m, f/7.8 Ritchey-Chr\'etien telescope and an Apogee U16M CCD ($4096 \times 4096$ 9-$\mu$m pixels, FoV: $26\arcmin \times 26\arcmin$). The telescope was autoguided, which allowed each star's center of light to remain on the same location of the CCD within about 3--4 pixels over the course of a night.

The 0.6-m Cassegrain photometric telescope at Rozhen, equipped with an FLI PL09000 CCD camera ($3056 \times 3056$ pixels, FoV: $17\arcmin \times 17\arcmin$) was used to observe two transits in 2010. 

An 0.4-m Schmidt-Cassegrain Meade LX200 GPS telescope at the Ankara University Observatory (Turkey) was employed to acquire two transit light curves. An Apoge ALTA-U47 CCD camera, ($1024 \times 1024$ pixels, FoV: $11\arcmin  \times 11\arcmin$) was used as a detector.

In addition, a transit on 2011 Oct 01 was taken with an 8-inch Schmidt-Cassegrain telescope and a G2-1600 CCD camera from Moravian Instruments Inc.\ (FoV: $48\arcmin \times 32\arcmin$) at a private amateur observatory in Herges-Hallenberg (Germany).

Data reduction was based on standard procedures including debiasing, dark correction, and flat-fielding using sky flats. Differential photometry was performed with respect to the comparison stars available in each field of view. The aperture size was optimized to achieve the smallest scatter in the resulting out-of-transit light curves. The light curves were detrended by fitting a second-order polynomial function of time along with a trial transit model,
using initial parameters obtained from the literature. This procedure was performed with the {\sc jktebop} code \citep{Southwortha, Southworthb}, which allows photometric trends to be modeled as polynomials up to 5th order. The best-fitting trend is subtracted from each light curve. Magnitudes are transformed into fluxes and normalised to have a mean of unity outside of the transit. The timestamps were converted to barycentric Julian dates in barycentric dynamical time \citep[BJD$_{\rm{TDB}}$,][]{Eastman}. To quantify the quality of each light curve, we used the photometric noise rate ($pnr$) defined as \begin{equation}
  pnr = \frac{rms}{\sqrt{\Gamma}}
\end{equation}
where the root mean square of the residuals, $rms$, is calculated from the light curve and a fitted model, and $\Gamma$ is the median number of exposures per minute \citep{Fulton}.

\subsection{Doppler measurements}\label{Sect.DopplerObs}

Seventeen precise RV measurements were acquired in 2010 and 2011 with the Hobby-Eberly Telescope \citep[HET,][]{Ramsey98} located in the McDonald Observatory (USA). The High Resolution Spectrograph \citep[HRS,][]{Tull98} was used with $R = 60000$ resolution. Data reduction was performed with the custom-developed ALICE code \citep{Nowaketal2013}.  It employs the standard iodine cell method and cross-correlation technique to calibrate data and measure the velocities. Using the I$_2$ cell method to measure RVs independently in 96-pixel long segments of our HET/HRS spectra, we obtained information about the imperfections in the initial Th-Ar dispersion curve and determine the instrumental profile. With this information, we cleaned the iodine lines from our spectra and constructed the cross-correlation function (CCF) from exactly the same parts of the spectra from which we measure RVs. For each epoch, the final value of the RV was taken to be the mean of the measurements from all 17 echelle orders. The measurement uncertainty was 28-–42 m~s$^{-1}$ at the 1-$\sigma$ level. The new RVs are listed in Table~\ref{Tab.RVs}.


\section{Results}\label{Results}

\subsection{Spectroscopic parameters for the host star}\label{Sect.StarParams}

To derive stellar properties, two spectra without iodine (acquired on 2010 May 29 and 2011 Aug 25) were combined to generate an averaged template spectrum.
This template has a signal-to-noise ratio of 300. The first step was to measure the stellar rotation velocity $v \sin I$ with the cross-correlation technique as described by \citet{Nowaketal2013}. The derived value of $v \sin I = 15.6\pm1.5$~km~s$^{-1}$ was used in further analyses.  Then, we measured equivalent widths of iron lines with the Automatic Routine for line Equivalent widths in stellar Spectra \citep[{\sc ARES},][]{Sousa07}. These results were used to determine the atmospheric parameters. The dataset was analysed with the {\sc TGVIT} code to determine the effective temperature $T_{\rm{eff}}$, surface gravity $\log g_{*}$, metallicity based on iron abundance [Fe/H], and micro turbulent velocity $v_{\rm{mic}}$. For a detailed description of the procedure,
we refer the reader to \citet{Takeda02, Takeda05}.

\begin{figure}
\epsscale{.50}
\plotone{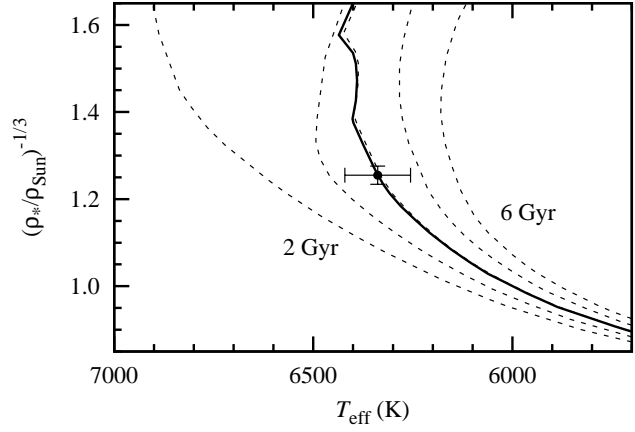}
\caption{Position of WASP-3 (a central dot) in the modified Hertzsprung-Russel diagram. The {\sc PARSEC} isochrones of the ages between 2 and 6 Gyr with a step of 1  Gyr are sketched with dashed lines. The best-fitting isochrone is drawn with a continuous line. Isochrones have been interpolated for redetermined iron abundance of $\rm{[Fe/H]}=-0.161$.\label{Fig.roT}}
\end{figure}

In the next step, the obtained parameters were used to determine the stellar mass $M_{*}$, luminosity $L_{*}$, and age. We employed {\sc PARSEC} isochrones in version 1.1 \citep{Bressan12} that were bilinearly interpolated to estimate the parameters and their uncertainties. The surface gravity is generally poorly constrained from spectroscopy \citep[see][and references therein]{Torres12}. For transiting planetary systems this parameter can be replaced by a mean stellar density $\rho_{*}$, which is determined more accurately from the transit light curve analysis (see Sect.~\ref{Sect.TrModel}). In this approach, the mean stellar density is known independently of theoretical stellar models. A mean stellar density based on the values of $\log g_{*}$ and $T_{\rm{eff}}$ from spectroscopy in conjunction with the evolutionary models is in agreement with the value derived from the transit photometry. The final stellar parameters are listed in Table~\ref{Tab.SpecResults}, and the position of WASP-3 in a modified Hertzsprung-Russell diagram is plotted in Fig.~\ref{Fig.roT}.

We also used alternative methods to check the spectroscopic parameters obtained above. We pursued the photometric approach of \citet{Adamow12}, which is based on the $BVJHK_{\rm{S}}$ photometry taken from the Tycho-2 catalogue \citep{Hog00} and 2MASS Point Source catalogue \citep{Cutri03}. The effective temperature of $6280\pm50$~K was obtained from an empirical calibration of six colours by \citet{Ramirez05}. A value of $\log g_{*} = 4.3$ was estimated from the empirical relation given by \citet{Straizys81}. A set of stellar parameters was also determined independently by spectrum modelling with the Spectroscopy Made Easy ({\sc SME}) code \citep{Valenti96}. The effective temperature was found to be 6440~K, $\log g_{*}= 4.3$, $\rm{[Fe/H]}=+0.01$, and $v_{\rm{mic}}=1.4$~km~s$^{-1}$.  Both methods gave results consistent with the values obtained with {\sc TGVIT}.

The stellar parameters we have determined are in a perfect agreement with the values reported by \citet{Torres12} and \citet{Pollacco08}, although we find a slightly less massive and less metal-rich host star ($M_{*}=1.11^{+0.08}_{-0.06}$~$M_{\odot}$, $\rm{[Fe/H]}=-0.161\pm0.063$). The effective temperature and surface gravity we derived ($T_{\rm{eff}}=6338\pm83$ K, $\log g_{*} =4.255^{+0.040}_{-0.037}$ in cgs units) deviate only from values reported by \citet{Montalto12}. Those literature values are higher, although were obtained by reanalysis of archival spectra from \citet{Pollacco08}. Differences are probably caused by different methodologies. The lithium abundance of $A(\rm{Li})=2.65\pm0.08$ dex, measured from the \ion{Li}{1} doublet at 6708 \AA, is in a range between 2.0 and 2.5 dex reported by \citet{Pollacco08}. It gives the system's age between 1.5 and 4 Gyr, based on an empirical relation obtained for open clusters \citep{Sestito05}. This estimate is consistent with the value of $3.9^{+1.3}_{-1.2}$ Gyr from isochrone fitting.


\subsection{Stellar activity}\label{Sect.Activity}

\begin{figure}
\epsscale{.50}
\plotone{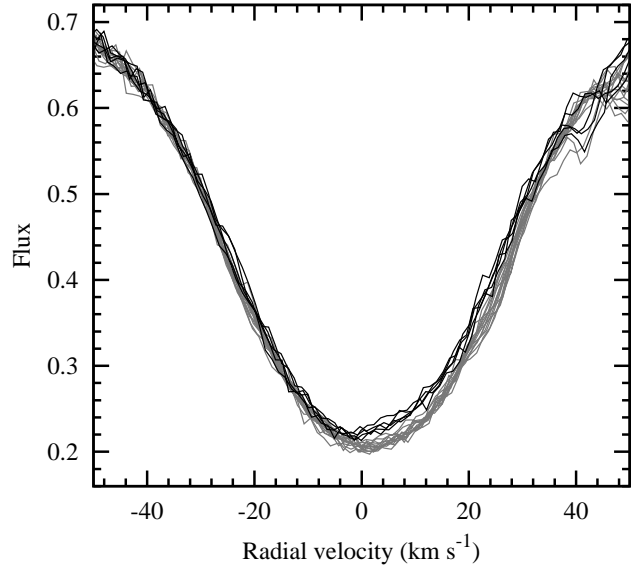}
\caption{Stack of individual spectra centred on the H$\alpha$ line and transformed into the RV domain. The spectra in a higher activity state are sketched with black lines, and these at the lower activity are drawn with grey lines. The profile becomes noticeable shallower and apparently blue shifted when the star is in the more active state.\label{Fig.Halfa}}
\end{figure}

The typical spectroscopic stellar activity indicators, such as Ca~II H (3968.47~{\AA}) and K (3933.66~{\AA}) lines and the infrared \ion{Ca}{2} triplet lines at 8498-8542~{\AA}, are outside of the wavelength range of our spectra.  Some alternative indicators, such as \ion{Na}{1} D1 (5895.92~{\AA}), \ion{Na}{1} D2 (5889.95~{\AA}), as well as \ion{He}{1} D3 (5875.62~{\AA}) lines, are seriously contaminated by the I$_{2}$ lines. Therefore, we used the H$\alpha$ (6562.808~{\AA}) line as a chromospheric activity indicator. The variation in the shape of the H$\alpha$ line between epochs can be seen in Fig.~\ref{Fig.Halfa}. To create this figure the wavelengths based on the ThAr comparison lamp were transformed into radial velocities, after correcting for the barycentric Earth motion (using a procedure of \citet{1980AAS...41....1S}), the absolute radial velocity of the WASP-3, and the RV variation produced by the transiting planet (see Sect.~\ref{Sect.RV}).

\begin{figure}
\epsscale{.50}
\plotone{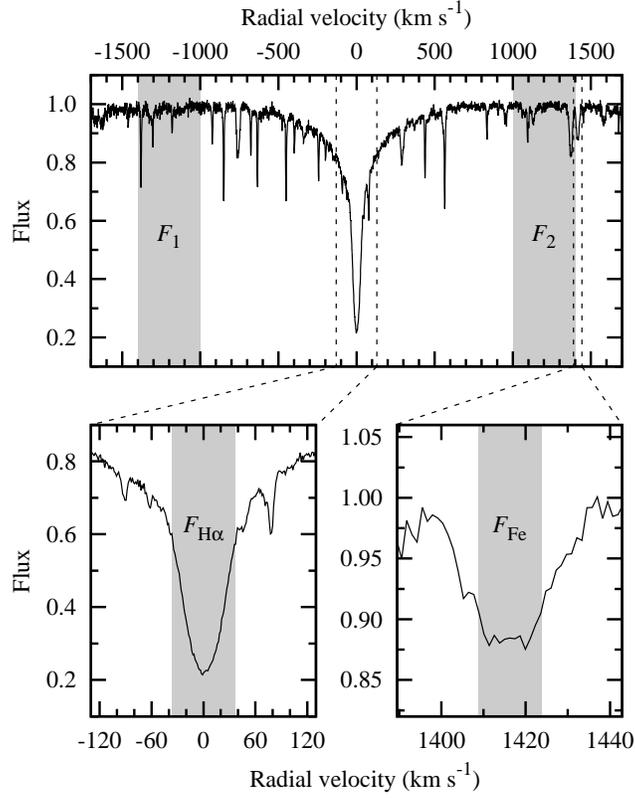}
\caption{Illustration of bands (greyed areas) used to determine the activity indices for the H$\alpha$ and control \ion{Fe}{1} 6593.884~{\AA} lines. Bottom panels zoom in on bands centred on the investigated lines. The spectrum is a template for RV measurements (i.e.\ without I${_2}$ lines) acquired on 2010 May 29.\label{Fig.acttml}}
\end{figure}

The lines that were observed at epochs of higher activity are slightly shallower and blue-shifted than the lines observed at epochs of lower activity. 
To quantify the stellar activity, we followed \citet{2012AA...541A...9G} and \citet[][and references therein]{2013ApJ...764....3R}. We determined the H$\alpha$ index ($I_{\rm{H\alpha}}$) as the ratio of the summed flux within a band of a width of 73~km~s$^{-1}$ ($\sim$1.6~{\AA}) centred on the core of the H$\alpha$ line ($F_{\rm{H\alpha}}$), and the summed flux within two reference bands ($F_{1}+F_{2}$) on both sides of the H$\alpha$ (between $-1400$ and $-1000$~km~s$^{-1}$ for $F_{1}$ and 1000 and 1400~km~s$^{-1}$ for $F_{2}$)
\begin{equation}
I_{\rm{H}\alpha} = \frac{F_{\rm{H}\alpha}}{F_{1}+F_{2}} \ .
\end{equation}
The adopted bands are illustrated in Fig.~\ref{Fig.acttml}. Values of the activity index were determined for 19 epochs, 17 listed in Table~\ref{Tab.RVs} and two template spectra taken in 2010 and 2011 that were used to determine precise RV measurements and stellar parameters. The uncertainties of $I_{\rm{H\alpha}}$ were calculated adopting Eq.\ (2) of \citet{2013ApJ...764....3R} 
\begin{equation}
\sigma_{I_{\rm{H}\alpha}} = I_{\rm{H}\alpha} \left( \left(\frac{\sigma_{\rm{F}_{\rm{H}\alpha}}}{F_{\rm{H}\alpha}}\right)^{2} + \left(\frac{\sqrt{\sigma^{2}_{\rm{F}_{1}}+\sigma^{2}_{\rm{F}_{2}}}}{F_{1}+F_{2}}\right)^{2} \right)^{1/2} \ ,
\end{equation}
where $\sigma_{\rm{F}_{\rm{H}\alpha}}$ is the $rms$ scatter in the continuum in the 0.5~{\AA} adjacent to the investigated line multiplied by the square root of the number of pixels in the line, and $\sigma_{\rm{F}_{1}}$ and $\sigma_{\rm{F}_{2}}$ are the values of the $rms$ scatter in the reference bands multiplied by the square root of the number of pixels in these bands.

To take possible instrumental effects into account, we also measured the index for the \ion{Fe}{1} 6593.884~{\AA} line \citep{2012AA...544A.125M}, which is expected to be insensitive to stellar activity. We used a band of a width of 14.5~km~s$^{-1}$ ($0.32$~{\AA}), centred on the core of the line, and the same reference bands (see Fig.~\ref{Fig.acttml}). As our stellar spectra were taken through the I$_{2}$ cell to precisely measure RVs, the wavelength regime relevant to our line indices may still contain weak I$_2$~lines (up to $\sim$4\% relative to the continuum). Thus, we measured the H$\alpha$ and \ion{Fe}{1} indices for the iodine flat-field spectra. These indices show the $rms$ variation at a level of 0.22\% and 0.21\% for the H$\alpha$ and \ion{Fe}{1} index, respectively. This intrinsic scatter is much smaller than the $rms$ scatter for the stellar $I_{\rm{H}\alpha}$ (2.39\%) and $I_{\rm{Fe}}$ (0.58\%). These findings show that presence of the weak I$_2$ lines in the combined stellar-iodine spectra do not introduce significant errors into determinations of the activity indices.

As most of our HET/HRS observations were obtained during ``priority 4'' time (non-ideal moon phase and weather), we also searched for any variability in water vapour lines near the H$\alpha$ wavelength regime. We measured the H$\alpha$ and \ion{Fe}{1} indices for very rapidly rotating and therefore line-depleted stars observed within the framework of the Pennsylvania-Toru\'n Search for Planets project \citep{2011IAUS..276..445N}. Using 203 spectra of these stars, acquired between January 2004 and August 2012, we found no seasonal (annual) variability of the telluric H$\alpha$ and \ion{Fe}{1} indices. Therefore, we conclude that neglecting the contribution of the telluric lines to the WASP-3 spectra does not introduce systematic errors in the determination of H$\alpha$ and \ion{Fe}{1} indices.

\begin{figure}
\epsscale{.50}
\plotone{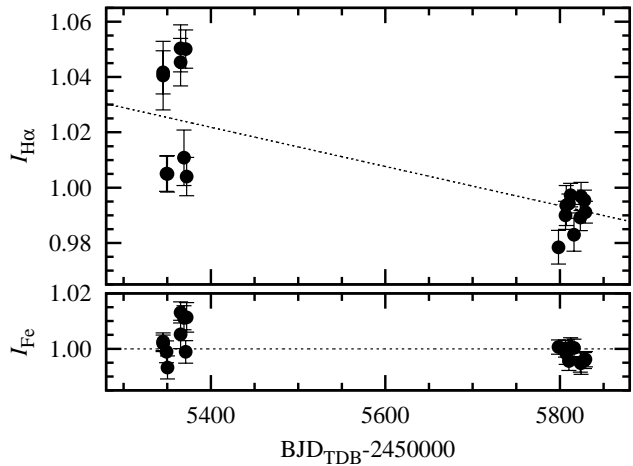}
\caption{Activity indices determined for H$\alpha$ (upper panel) and \ion{Fe}{1} (bottom panel) lines as a function of time, normalized to the mean values for ease of comparison. Measurements grouped around 2455350 and 2455800 BJD were obtained in 2010 and 2011, respectively.\label{Fig.act}}
\end{figure}

Figure \ref{Fig.act} shows the behaviour of $I_{\rm{H}\alpha}$ and $I_{\rm{Fe}}$ (both normalized to their mean values) as a function of time. While $I_{\rm{Fe}}$ remains constant between the 2010 and 2011 observing seasons, there is a noticeable decrease of $I_{\rm{H}\alpha}$. A linear regression gives a gradient of $\Delta I_{\rm{H}\alpha}=(-2.3\pm0.5)\times10^{-6}$~d$^{-1}$. Moreover, measurements in 2010 exhibit larger scatter, compared to the less active stage in 2011. This effect could be caused by slowly-evolving active regions on the stellar surface, which are modulated by stellar rotation.


\subsection{Transit model}\label{Sect.TrModel}

\begin{figure}
\epsscale{1.0}
\plotone{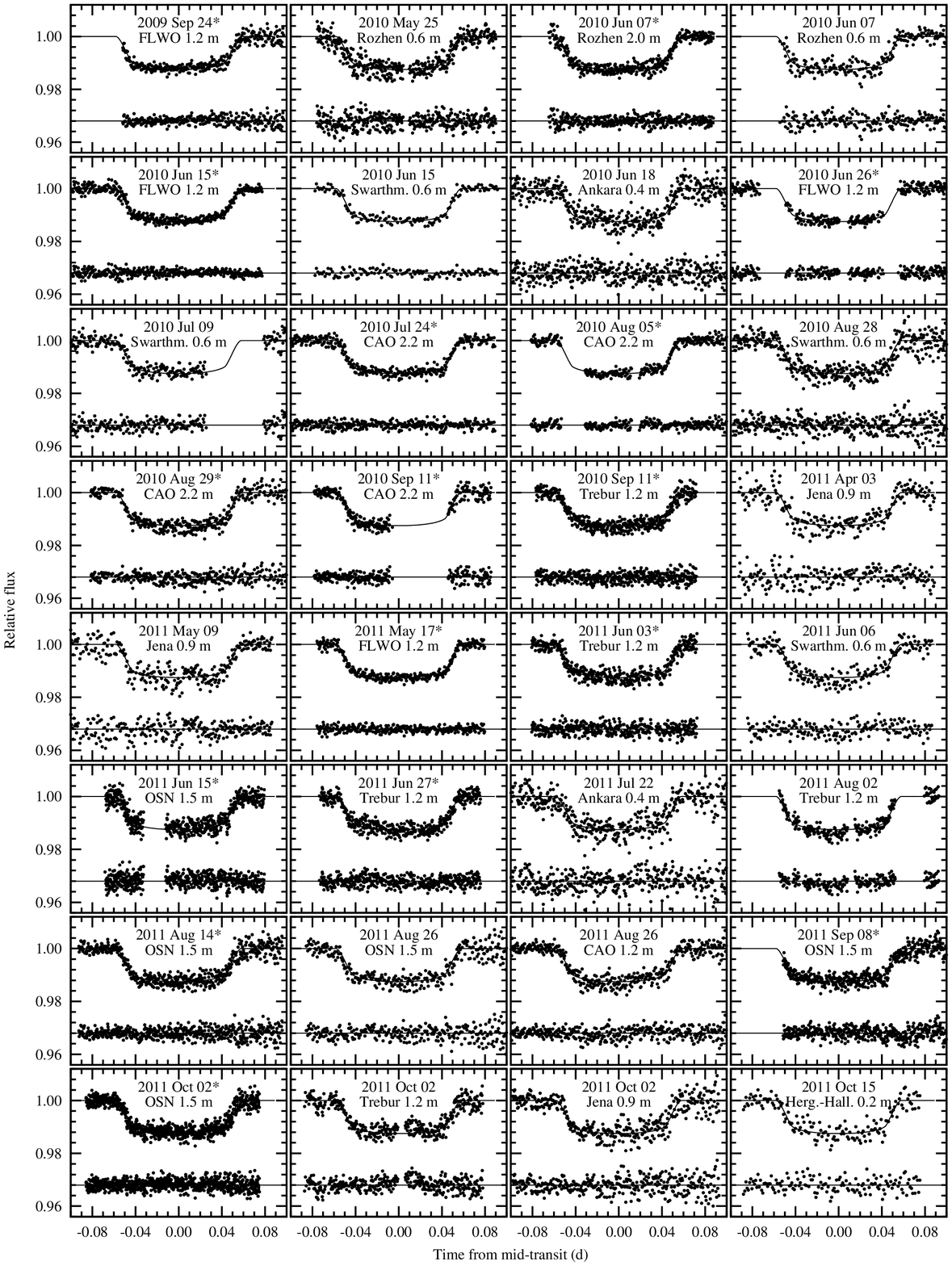}
\caption{Transit light curves obtained for WASP-3~b. Three-point binning was applied for light curves acquired at the Ankara University Observatory for clarity. Light curves which were used to produce a transit template are marked with an asterisk in the date of observation. \label{Fig.LCs}}
\end{figure}

We selected a collection of the highest-quality light curves for modelling with the Transit Analysis Package\footnote{http://ifa.hawaii.edu/users/zgazak/IfA/TAP.html} \citep[{\sc TAP} v2.1,][]{Gazak12} to obtain transit parameters. The selection was done iteratively. Light curves were sorted according to the photometric noise rate ($pnr$, see Sect.~\ref{Sect.TransitObs}). The fitting procedure started with a few light curves with the smallest value of $pnr$ and subsequent light curves were added in next iterations. We noticed that including datasets with $pnr>1.64$ mmag degraded the quality of the fit, so the procedure was interrupted, and finally a set of sixteen best-quality light curves (indicated in Table~\ref{Tab.LCObservations} and Fig.~\ref{Fig.LCs}) was used to generate the transit model. Despite of its relatively good value of $pnr = 1.43$ mmag, a light curve observed on 2011 Aug 02 was excluded because it is incomplete and exhibits the correlated noise which can be clearly seen in the residuals (Fig.~\ref{Fig.LCs}). The final sample comprises eleven light curves in $R$ band, two in $r'$, two in $i'$, and one in $I$. {\sc TAP} uses the Markov Chain Monte Carlo (MCMC) method, with the Metropolis-Hastings algorithm and a Gibbs sampler, to find the best-fitting parameters based on the transit model of \citet{MandelAgol}. In estimating the parameter uncertainties, the wavelet-based technique of \citet{Carter09} is used to take into account time-correlated noise. It has been shown that this approach provides the most reliable parameters and error estimates \citep[e.g.][]{Hoyer12}.

The {\sc TAP} code employs the quadratic limb-darkening (LD) law. As the initial values for the fitting procedure, we used theoretical values of LD coefficients (LDCs) from tables of \citet{Claret11}, linearly interpolated with the \textsc{EXOFAST} applet\footnote{http://astroutils.astronomy.ohio-state.edu/exofast/limbdark.shtml} \citep{Eastman12} for the WASP-3 stellar parameters that were presented in Sect.~\ref{Sect.StarParams}. In the final iteration, the linear LDCs were allowed to vary freely. The quadratic terms were allowed to vary subject to Gaussian priors centered on the theoretical values, with a Gaussian width of 0.05. This approach is justified when data are not precise enough to solve for both the linear and quadratic LDCs. When both coefficients were allowed to vary, the fitting procedure sometimes gave unphysical results. We also considered a scenario in which the LDCs were held fixed at the theoretical values. This approach does not take into account any uncertainty in the LDCs, and the resulting parameter uncertainties were correspondingly reduced by up to 12\%.

Three of the model parameters---the orbital inclination $i_{\rm{b}}$, semimajor-axis scaled by stellar radius $a_{\rm{b}}/R_{*}$, and planetary to stellar radii ratio $R_{\rm{b}}/R_{*}$---were required to be consistent across all of the light curves. The LDCs were also required to be the same for all of the data in a given bandpass. The orbital period was fixed at a value of 1.8468349~d, taken from \citet{Nascimbeni13}, and the mid-transit times of individual light curves were left as free parameters to account for possible timing variations. In addition, the fitting procedure accounted for the uncertainties in the linear trends in individual datasets that were removed at the preprocessing stage (Sect.~\ref{Sect.TransitObs}).
Since the RVs are consistent with a circular orbit (Sect.~\ref{Sect.RV}), we assumed in the transit analysis that the orbit of WASP-3~b is perfectly circular.

Ten MCMC chains, each containing $10^6$ steps, were computed. The individual chains were combined to get final posteriori probability distributions. The first 10\% of the links in each chain were discarded before calculating the best-fitting parameter values and their uncertainties. They were determined by taking the median value of marginalised posteriori probability distributions, which were found to be unimodal. The 15.9 and 84.1 percentile values of the cumulative distributions were used to define the upper and lower 1\,$\sigma$ uncertainties.

\begin{figure}
\epsscale{.50}
\plotone{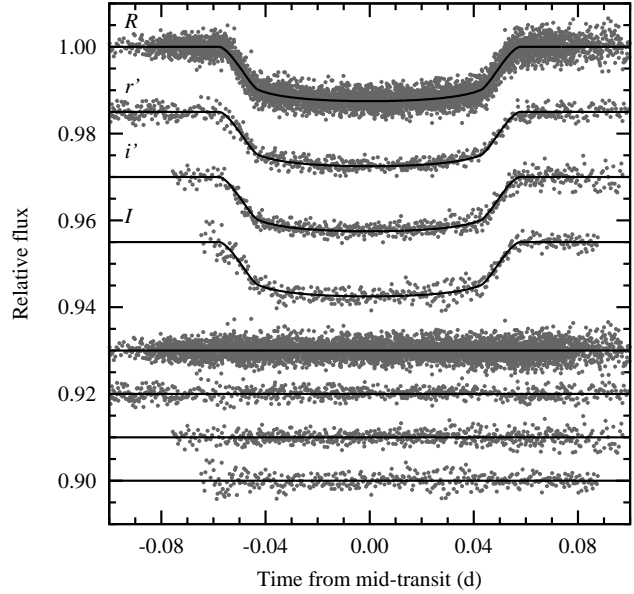}
\caption{Composite light curves in individual filters with the best-fitting transit models. The residuals are plotted in the bottom.\label{Fig.TrModel}}
\end{figure}

Table~\ref{Tab.TranResults} gives the results, and compares them to other determinations in the literature. The optimized transit models for different filters are plotted in Fig.~\ref{Fig.TrModel}. The values reported in this work agree with most of the previous determinations. Some of the comparisons are not straightforward because different methods have been used for parameter estimation; in particular many of the previous determinations did not take time-correlated noise into account, and the reported uncertainties are likely to have been underestimated.  Nevertheless, our more conservative determinations are generally more precise than most of those from previous studies. The linear LDCs of $u_{\rm{R}}=0.24\pm0.04$, $u_{r'}=0.28\pm0.06$, $u_{i'}=0.18\pm0.06$ were found to be systematically smaller than the theoretical values ($u^{\rm{t}}_{\rm{R}}=0.28$, $u^{\rm{t}}_{r'}=0.30$, $u^{\rm{t}}_{i'}=0.23$), but within the uncertainties. The exception is $u_{\rm{I}}=0.23\pm0.10$ which was found to be slightly greater than the theoretical value of $u^{\rm{t}}_{\rm{I}}=0.21$. This finding seems not to be conclusive because it is based on a single light curve which could be affected by imperfect detrending (there were few observations before the beginning of the transit). Interestingly, \citet{Nascimbeni13}, who fitted a linear LDC in $R$ band, also obtained a smaller value of $u_{\rm{R}}$. These subtle differences between observed and theoretical values may be caused by imperfect LD tables or presence of active areas on the stellar surface \citep{Csizmadia12}. They could also be caused by biases related to transit fitting; for example the LDCs determined from transit light curves have been found to depend on the transit parameter \citep{Howarth11}.

The fitting procedure was repeated with $i_{\rm{b}}$, $a_{\rm{b}}/R_{*}$, and $R_{\rm{b}}/R_{*}$ allowed to vary between individual epochs and individual filters to search for variations in these parameters with time and bands. The linear LDCs were allowed to vary around the values derived earlier, subject to a Gaussian prior defined by the previously derived uncertainties. We found that none of the parameters show a periodic modulation or a long-timescale trend. These results cast into doubt the transit duration variation postulated by \citet{Eibe12}. Possible variations in $R_{\rm{b}}/R_{*}$ could be induced by stellar activity if the fraction of the stellar surface covered by spots changes from transit to transit \citep[e.g.][]{Carter11}. In addition, occultations of dark spots by a planetary disc would produce apparent brightening in transit light curves \citep{Schneider00,Silva03}, which are not seen in any of the highest quality datasets.  There is therefore no purely photometric evidence for stellar activity. Moreover, no significant differences in $R_{\rm{b}}/R_{*}$ have been found between $R$, $I$, $r'$, and $i'$ filters.


\subsection{Mid-transit times}\label{Sect.TrTimes}

\begin{figure}
\epsscale{1.0}
\plotone{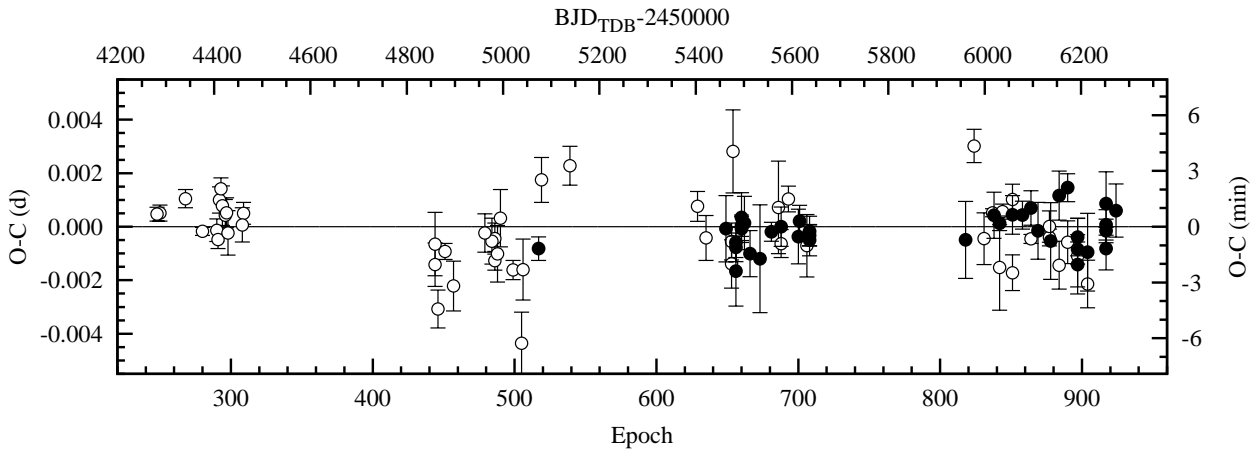}
\caption{O--C diagram for transit timing of WASP-3~b. Open and filled symbols mark times from the literature and our new results, respectively. \label{Fig.OC}}
\end{figure}

The transit model based on the best-quality data (Sect.~\ref{Sect.TrModel}) was used as a template to determine the mid-transit times for each individual light curve, using the TAP code. The parameters $i_{\rm{b}}$, $a_{\rm{b}}/R_{*}$, $R_{\rm{b}}/R_{*}$, and the LDCs were allowed to vary, subject to Gaussian priors based on the results described previously. This approach guarantees that the unceratinties in the model parameters are taken into account in the error budget for each mid-transit time. For each individual light curve fit, the orbital period is of little consequence but for completeness it was held fixed as in Sect.~\ref{Sect.TrModel}. The mid-transit time, as well as the flux slope and intercept, were taken to be free parameters. The MCMC analysis used ten chains of a length of $10^5$ steps for each light curve.  Five transits were observed with more than one telescope. In such cases the light curves were fitted simultaneously to increase timing precision by up to 31\%, depending on the quality of individual datasets. The results for the mid-transit times are listed in Table~\ref{Tab.Timing}. They were combined with 53 published mid-transit times to refine the transit ephemeris. The mid-transit time from the discovery paper \citep{Pollacco08} was excluded because its value was determined as an average from various datasets. We used redetermined times by \citet{Nascimbeni13} who performed a uniform analysis of all datasets available to those authors. We also used times reported by \citet{Eibe12} and \citet{Montalto12}. As a result of a linear fit which uses individual timing errors as weights, we obtained the orbital period of $P_{\rm{b}}=1.8468351\pm0.0000004$~d and the time of transit at cycle zero of $T_{0}=2454143.85112\pm0.00024$ BJD$_{\rm{TDB}}$ with reduced $\chi^2=3.3$. We adopted the cycle numbering starting from the ephemeris given by \citet{Pollacco08}. The O--C (observed minus calculated) diagram for transit timing is plotted in Fig.~\ref{Fig.OC}.

The timing residuals (the observed mid-transit times after subtracting the best-fitting linear function of epoch number) were searched for any periodic variation using a Lomb-Scargle periodogram \citep{Lomb,Scargle}. The strongest peak was found to be insignificant, with a false alarm probability (FAP) equal to 52\%. This value was determined empirically by a bootstrap resampling method which generates $10^5$ datasets with the randomly permuted O--C values at the original observing epochs, and determines the fraction of resampled datasets with power higher than the original dataset.

The value of reduced $\chi^2$ for the linear ephemeris is far from unity.  In principle this may be caused by a quasi-periodic or non-periodic (chaotic) TTV signal, or a long-timescale TTV signal.
The first scenario is doubtful because the new observations produce no significant peak in the periodogram of timing residuals. The detection of the putative TTV signal by \citet{Maciejewski10} was probably caused by small-number statistics. The second possibility was pointed out by \citet{Nascimbeni13} and it could be generated by a specific two-planet configurations close to mean-motion resonances or by configurations with more than one perturbing body. The third scenario, employing a parabolic fit reflecting any secular variation in the orbital period, has already been ruled out by \citet{Montalto12}. The high value of $\chi^2$ could also be a simple consequence of underestimating the uncertainties in the mid-transit times. It has been shown that Monte Carlo, bootstrapping, or residual-shift (prayer-bead) methods may lead to underestimated uncertainties by a factor of up to four \citep[see e.g.][]{Maciejewski13}. The wavelet-based techniques that are implemented in TAP allow one to take into account time-correlated noise in the photometric data, and seem to provide the most reliable uncertainty estimates \citep{Carter09}. Transit timing may also be affected by systematic effects caused by weather conditions (e.g. passing thin clouds, variable atmospheric extinction), instrumental factors (e.g. imperfect autoguiding, variable characteristic of the CCD matrix), or data reduction (e.g. detrending and normalization of a light curve). These effects are difficult to account for in the error budget, and may generate outliers in the O--C diagram. If the sample of transit times is limited to these reported in this paper and two points reported by \citet{Gibson08} (taken for a longer timespan), the reduced $\chi^2$ for a linear ephemeris is equal to 1.07. This result shows that our transit times are consistent with the linear ephemeris. We also examined those light curves which are the sources of mid-transit times lying more than 1~$\sigma$ away from zero in the O--C diagram (note that no point deviates by more than 3 $\sigma$). Most of these light curves have incomplete coverage of a transit or were obtained on nights with variable conditions (thin clouds, deteriorating transparency, or a high airmass range). Thus, we conclude that the large scatter in the O--C diagram is likely a consequence of underestimated uncertainties due to observational and/or data-analysis factors.


\subsection{Orbital fit}\label{Sect.RV}

The Systemic Console software \citep{Meschiari09} was used to refine the orbital parameters of WASP-3~b. The data from \citet{Pollacco08}, \citet{Tripathi10}, and \citet{Simpson10} were combined with our new RV measurements to derive the planet's minimum planetary mass $M_{\rm{b}} \sin i_{\rm{b}}$ and semi-major axis $a_{\rm{b}}$. \citet{Pollacco08} and \citet{Simpson10} reported RV measurements obtained with the SOPHIE spectrograph on the 1.9-m telescope at the Observatoire de Haute Provence. \citet{Tripathi10} used the High Resolution Echelle Spectrometer (HIRES) on the Keck I telescope at the W.~M.\ Keck Observatory on Mauna Kea to study the RM effect and to refine the orbital parameters. The data points that were obtained during transits ($\pm80$ min from the expected mid-transit time, based on the transit ephemeris refined in Sect.~\ref{Sect.TrTimes}) were removed from the sample because our orbital model does not account for the RM effect. In the fitting procedure, the RV offsets between individual instruments were allowed to vary to account for differences in calibration of system velocities. The mid-transit times after 3-$\sigma$ clipping were included in the RV model to better constrain the mean anomaly at an initial epoch. The orbital period was allowed to be a free parameter, in order to verify the value obtained from transit timing alone, and to take
into account the uncertainty in the orbital period
when computing the uncertainties in the other parameters.
For this analysis we adopted the value of $M_{*}$ with uncertainties obtained in Sect.~\ref{Sect.StarParams}.

\begin{figure}
\epsscale{0.5}
\plotone{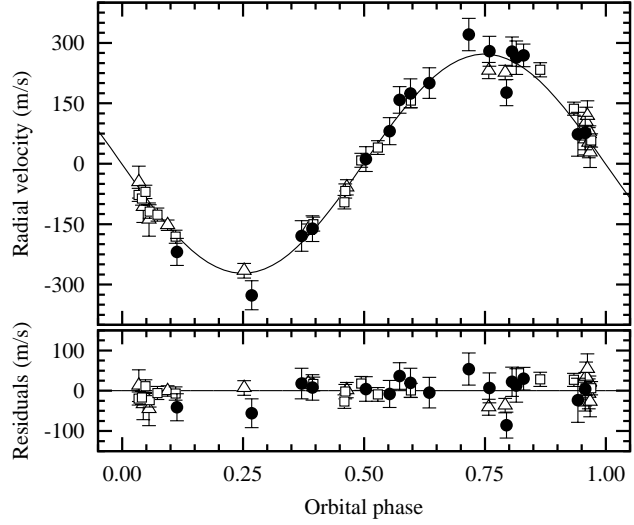}
\caption{Phase-folded RV variations induced by the WASP-3 planet. Open triangles denote measurements from \citet{Pollacco08} and \citet{Simpson10}. Open squares come from \citet{Tripathi10}. Our new data points are marked with filled circles. The best-fitting model assuming the circular orbit is sketched with a continuous line. \label{Fig.W3RV}}
\end{figure}

The Nelder-Mead minimization algorithm was used to find the best-fitting Keplerian orbit solution. The MCMC method was used to determine parameter uncertainties. The MCMC chain was $10^6$ steps long, and the first 10\% configurations were discarded. The scale parameters were set empirically in a series of attempts to get the acceptance rate of the MCMC procedure close to the optimal value of $0.25$. For each parameter, the standard deviation was taken as the final error estimate. Two scenarios with a circular and eccentric orbits were considered. In the latter case, the eccentricity was found to be $e_{\rm{b}}=0.02\pm0.01$ with reduced $\chi^2$ equal to 1.18 and $rms_{\rm{rv}}$ equal to 26.7~m~s$^{-1}$. The circular-orbit model
fits nearly as well as the eccentric-orbit model,
with reduced $\chi^2=1.26$ and $rms_{\rm{rv}}=27.5$~m~s$^{-1}$. As the significance of the non-zero eccentricity is low and there is no significant improvement in RV residuals, we discard the non-circular solution and adopt $e_{\rm{b}}=0.0$ in subsequent calculations. This approach is also supported by observations of planetary occultations at the time expected
for a circular orbit \citep{Zhao12}. The orbital solution is illustrated in Fig.~\ref{Fig.W3RV} and the redetermined parameters are given in Table~\ref{Tab.OrbiResults}.

Our RV dataset and that of \citet{Tripathi10} contain measurements spanning two consecutive observing seasons. Splitting these datasets into individual seasons and keeping relative offsets as free parameters could reveal possible long-term RV shifts caused by instrumental effects, small-number statistics, stellar activity, or additional bodies on wide orbits. This approach results in the value of $rms_{\rm{rv}}$ reduced by 10\%. However, the relative offsets were found to be $6\pm16$ and $40\pm27$ m~s$^{-1}$ for the dataset of \citet{Tripathi10} and our new measurements, respectively. Both values are consistent with zero within $\sim$$1.5\sigma$. Therefore there is no compelling evidence for any RV trends over the timespan of 1 year.


\subsection{Constraints on additional planet}\label{Sect.Constraints}

The absence of any detectable periodic TTV signal, and the absence of any RV evidence for a departure from a single Keplerian orbit, allows us to place constraints on the properties of any hypothetical second planet in the system. The {\sc Mercury 6} package \citep{Chambers99} with the Bulirsch--Stoer integrator was used to generate a set of synthetic O--C diagrams for WASP-3~b in the presence of a fictitious perturbing planet. The mass of the fictitious planet was set at 0.5, 1, 5, 10, 50, 100, and 500 $M_{\rm{Earth}}$ (Earth masses), and the initial orbital distance varied between 0.006 and 0.2 AU with a step of $2\times10^{-6}$ AU. The initial orbital longitude of WASP-3~b was set at a value calculated for cycle zero, and the initial longitude of the fictitious planet was shifted by $180\degr$. The system was assumed to be coplanar, with both orbits initially circular.  The integration time covered 1250 days, i.e., the time span of the transit observations. The value of $rms$ was calculated for each synthetic O--C diagram. Then, for each value of the orbital distance, we determined the range of planet masses for which the calculated $rms_{\rm{ttv}}$ was smaller than 80~s. Before calculating $rms_{\rm{ttv}}$, a 3-$\sigma$ clipping was applied to remove ``outlying'' data points. An upper mass of the fictitious planet at the detection limit was found by linear interpolation for masses below 500 $M_{\rm{Earth}}$. If $rms_{\rm{ttv}}$ was found to be generated by a more massive body, the limiting mass was extrapolated using a linear trend as fitted to 100 and 500 $M_{\rm{Earth}}$. Most of orbits located within $\sim$3.5 Hill radii of WASP-3~b (i.e., close to 1:1 orbital period commensurability) were found to be highly unstable and planetary close encounters or planet ejections occurred during the relatively short time of integration.

\begin{figure}
\epsscale{0.6}
\plotone{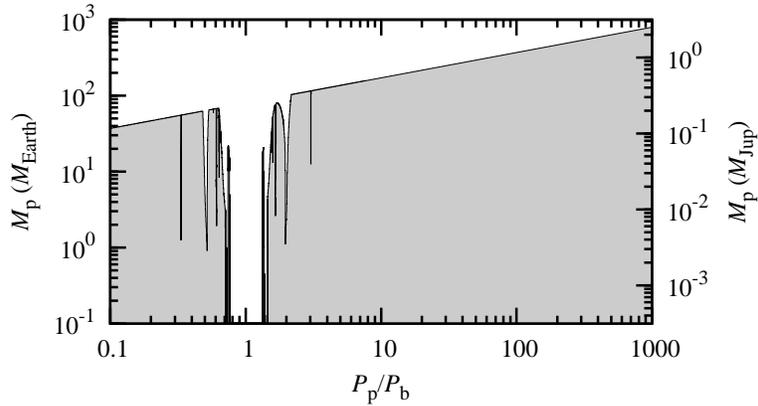}
\caption{Upper mass limit for a fictitious second planet in the WASP-3 system, based on timing and RV datasets, as a function of an orbital period of that planet, $P_{\rm{p}}$. The greyed area marks configurations which are below our detection threshold. \label{Fig.Limit}}
\end{figure}

A similar approach was applied to the RV dataset. The value of $rms_{\rm{rv}}$ (Sect.~\ref{Sect.RV}) was used to calculate the mass limit as a function of the semi-major axis of the fictitious planet. Again, circular orbits were assumed. Then, both criteria were combined to obtain the upper mass limit of the possible second planet, based on transit timing and RV datasets. The results are plotted in Fig.~\ref{Fig.Limit}. 

While the RV method gives tighter constraints for most configurations, the TTV technique is sensitive to low-mass perturbers close to low-order mean-motion resonances. The RV dataset limits masses of inner perturbers to $\sim$40 $M_{\rm{Earth}}$ for tightest orbits and to $\sim$70 $M_{\rm{Earth}}$ for orbits close to WASP-3~b. The transit timing constrains masses of fictitious planets down to 1.7, 0.9, and 1.9 $M_{\rm{Earth}}$ in inner 3:1, 2:1, and 5:3 orbital resonances, respectively. For the outer perturbers, the RV method limits their masses down to $\sim$100 $M_{\rm{Earth}}$ for the most close-in orbits. The TTV method allows us to probe masses down to 2.6, 0.8, and 13 $M_{\rm{Earth}}$ in outer 5:3, 2:1, and 3:1 orbital resonances, respectively.


\section{Summary}\label{Sect.Summary}

We have acquired 32 new transit light curves for the planet WASP-3~b, and 17 precise radial velocity measurements for the WASP-3 host star. Our new data cover a timespan of 2 years from 2009 to 2011. The tangible result of our study is refining stellar, orbital, and planetary parameters with improved precision. Our studies of the stellar activity of WASP-3 confirm its long timescale variation reported by \citet{Montalto12} and also reveal a night-to-night variability when the star was in a more active state. These short timescale variations are likely to be caused by active regions that are carried around by stellar rotation.

Our result for the planetary mass ($M_{\rm{b}} = 1.77^{+0.11}_{-0.09}$~$M_{\rm{Jup}}$) agrees with the value reported by \citet{Pollacco08}, and the radius ($R_{\rm{b}} = 1.346 \pm 0.063$~$R_{\rm{Jup}}$) falls between estimates of \citet{Gibson08} and \citet{Christiansen11}. Additional RV measurements provide tighter constraints on the orbital eccentricity ($e_{\rm{b}}=0.02\pm0.01$) than \citet{Pollacco08}. The orbit of WASP-3~b is expected to be circular because its circularization timescale of 1-14 Myr for the tidal dissipation parameter $Q_{\rm{p}}$ between $10^5$ and $10^6$ is much shorter than the system's age of $3.9^{+1.3}_{-1.2}$ Gyr.

Despite all of this observational effort, no evidence for the presence of the additional planet in the WASP-3 system was found. Published hints for both periodic and chaotic variations in transit timing are likely caused by underestimated uncertainties and systematic effects affecting photometric measurements. We find a spectroscopic sign of variation in stellar activity for WASP-3 that is reported by \citet{Montalto12}. However, our high-precision photometry shows no evidence for starspot-crossing anomalies or other effects that stellar activity might have on transit light curves. The current precision of transit timing observations allows us to rule out Earth-mass planetary companions of WASP-3~b near the lowest-order mean-motion resonances. The radial-velocity data show no sign of additional bodies, and in particular no long-term trend over a few years. We note, however, that the portion of parameter space for additional bodies that remains unexplored is still significant.

Analysis of a sample of hot Jupiter candidates observed with the {\it Kepler} space telescope \citep{Borucki} shows that the overwhelming majority of these planets are devoid of close planetary companions \citep{Latham11,Steffen12}. This effect is interpreted as a result of the dynamical evolution of planetary systems containing close-in giant planets. In this context, a lack of confirmation of TTVs for WASP-3~b is consistent with expectations arising from the {\it Kepler} survey.



\acknowledgments

GM and GN acknowledge the financial support from the Polish Ministry of Science and Higher Education through the Iuventus Plus grants IP2010 023070 and IP2011 031971. AN, GN, and MA are supported by the Polish Ministry of Science and Higher Education grant N N203 510938. GN is also supported by the Faculty of Physics, Astronomy and Informatics grant no. 1627-A. MA is also supported by the Polish National Science Centre grant no.\ UMO-2012/05/N/ST9/03836.
AW is supported by the NASA grant NNX09AB36G. RN would like to acknowledge financial support from the Thuringian
government (B~515-07010) for the STK CCD camera used in this project.
RN, RE, SR, and CA would like to thank the German National Science
Foundation (Deutsche Forschungsgemeinschaft, DFG) for support
in the Priority Programme SPP 1385 on the {\em First ten Million
years of the Solar System} in projects NE 515/34-1 \& -2,
NE 515/33-1 \& -2, and NE 515/35-1 \& -2.
JNW and MJH gratefully acknowledge support from the NASA Origins program
and the MIT UROP office.
DWL acknowledge partial support from the Kepler mission
under NASA Grant and Cooperative Agreement NCC2-1390 with the Smithsonian
Astrophysical Observatory. UK would like to thank DFG for support in SPP 1385 in
project number RE 882/12-1. TOBS and CG would like to thank
DFG for support in project NE 515/30-1.
MS would like to thank DFG for support in project NE 515/36-1.
CG and MM would like to thank DFG for support in project MU 13-1.
We would like to thank DFG for travel support to
Calar Alto runs in projects NE 515/40-1.
DD acknowledges the financial support of the projects DO 02-362 and DDVU 02/40-2010 of the Bulgarian National Science Fund, observing grant support from the Institute of Astronomy and NAO, Bulgarian Academy of Sciences, and travel funds from the PAN/BAN exchange and joint research project ``Spectral and photometric studies of variable stars''. 
We thank the HET resident astronomers and telescope operators for continuous support.
The HET is a joint project of the University of Texas
at Austin, the Pennsylvania State University, Stanford
University, Ludwig-Maximilians-Universit\"at M\"unchen,
and Georg-August-Universit\"at G\"ottingen. The HET is
named in honor of its principal benefactors, William P.
Hobby and Robert E. Eberly. The Center for Exoplanets
and Habitable Worlds is supported by the Pennsylvania
State University, the Eberly College of Science, and the
Pennsylvania Space Grant Consortium.

{\it Facilities:}  \facility{HET}, \facility{CAO:2.2m}, \facility{FLWO:1.2m}, \facility{CAO:1.2m}.



\clearpage

\clearpage

\begin{deluxetable}{rlrlcclccc}
\tabletypesize{\scriptsize}
\tablecaption{Observed transits and light-curve features.\label{Tab.LCObservations}}
\tablewidth{0pt}
\tablehead{
\colhead{\#} & \colhead{Date UT} & \colhead{$E$} & \colhead{Telescope} & \colhead{Filter} & \colhead{Airmass} & \colhead{Weather} & \colhead{$\Gamma$} & \colhead{$pnr$} & \colhead{Templ.} 
}
\startdata
 1 & 2009 Sep 24 & 517 & FLWO 1.2 m & $i'$ & $1.02\rightarrow1.75$ & clear & 1.36 & 1.40 & $\surd$\\
 2 & 2010 May 25 & 649 & Rozhen 0.6 m & $I_{\rm{C}}$ & $1.39\rightarrow1.01$ & clear  & 1.82 & 1.94 & \\
 3 & 2010 Jun 07 & 656 & Rozhen 2.0 m & $I_{\rm{C}}$ & $1.58\rightarrow1.02$ & clear  & 1.77 & 1.24 & $\surd$ \\
    &                       &       & Rozhen 0.6 m & $I_{\rm{C}}$ & $1.51\rightarrow1.01$ & clear  & 0.95 & 2.35 & \\
 4 & 2010 Jun 15 & 660 & FLWO 1.2 m & $r'$ & $1.88\rightarrow1.00$ & clear  & 1.54 & 0.99 & $\surd$\\
    & 				 &       & Swarthm. 0.6 m & $I_{\rm{C}}$ & $1.06\rightarrow1.00\rightarrow1.15$  & clear & 0.52 & 1.77 & \\
 5 & 2010 Jun 18 & 662 & Ankara 0.4 m & $R_{\rm{C}}$ & $1.24\rightarrow1.00\rightarrow1.21$ & clear  & 4.60 & 2.52 & \\
 6 & 2010 Jun 26 & 666 & FLWO 1.2 m & $r'$ & $1.14\rightarrow1.00\rightarrow1.45$ & partly cloudy  & 1.36 & 1.15 & $\surd$ \\
 7 & 2010 Jul 09 & 673 & Swarthm. 0.6 m & $I_{\rm{C}}$ & $1.06\rightarrow1.00\rightarrow1.65$ & clear  & 1.02 & 2.09 & \\
 8 & 2010 Jul 24 & 681 & CAO 2.2 m & $R_{\rm{C}}$ & $1.00\rightarrow1.71$ & clear  & 1.28 & 1.12 & $\surd$ \\
 9 & 2010 Aug 05 & 688 & CAO 2.2 m & $R_{\rm{C}}$ & $1.00\rightarrow1.59$ & clear  & 1.22 & 1.14 & $\surd$ \\ 
10 & 2010 Aug 28 & 700 & Swarthm. 0.6 m & $I_{\rm{C}}$ & $1.00\rightarrow2.08$ & clear  & 1.36 & 2.79 & \\
11 & 2010 Aug 29 & 701 & CAO 2.2 m & $R_{\rm{C}}$ & $1.05\rightarrow2.91$  & cirrus clouds & 1.46 & 1.64 & $\surd$ \\ 
12 & 2010 Sep 11 & 708 & CAO 2.2 m & $R_{\rm{C}}$ & $1.01\rightarrow1.68$  & clear & 1.71 & 1.28 & $\surd$ \\
     &                      &        & Trebur 1.2 m & $R_{\rm{B}}$ & $1.08\rightarrow1.94$  & clear & 2.14 & 1.24 & $\surd$ \\
13 & 2011 Apr 03 & 818 &  Jena 0.9 m & $R_{\rm{B}}$ & $2.09\rightarrow1.08$  & clear & 0.83 & 3.39 & \\
14 & 2011 May 09 & 838 & Jena 0.9 m & $R_{\rm{B}}$ & $1.75\rightarrow1.04$  & clear & 1.05 & 2.86 & \\
15 & 2011 May 17 & 842 &  FLWO 1.2 m & $i'$ & $1.23\rightarrow1.00\rightarrow1.02$  & clear & 1.36 & 0.94 & $\surd$ \\
16 & 2011 Jun 03 & 851 & Trebur 1.2 m & $R_{\rm{B}}$ & $1.23\rightarrow1.03\rightarrow1.04$  & clear & 2.14 & 1.32 & $\surd$ \\
17 & 2011 Jun 06 & 869 & Swarthm. 0.6 m & $r'$ & $1.01\rightarrow1.00\rightarrow1.31$  & clear & 0.81 & 2.72 & \\
18 & 2011 Jun 15 & 858 & OSN 1.5 m & $R_{\rm{C}}$ & $1.62\rightarrow1.02$  & clear & 2.73 & 1.43 & $\surd$ \\
19 & 2011 Jun 27 & 864 & Trebur 1.2 m & $R_{\rm{B}}$ & $1.05\rightarrow1.03\rightarrow1.17$  & clear & 2.39 & 1.24 & $\surd$ \\
20 & 2011 Jul 22 & 878 & Ankara 0.4 m & $R_{\rm{C}}$ & $1.09\rightarrow1.00\rightarrow1.26$  & clear & 4.63 & 2.67 & \\
21 & 2011 Aug 02 & 884 & Trebur 1.2 m & $R_{\rm{B}}$ & $1.03\rightarrow1.38$  & partly cloudy & 1.82 & 1.43 &  \\
22 & 2011 Aug 14 & 890 & OSN 1.5 m & $R_{\rm{C}}$ & $1.04\rightarrow2.64$  & clear & 2.00 & 1.53 & $\surd$ \\
23 & 2011 Aug 26 & 897 & OSN 1.5 m & $R_{\rm{C}}$ & $1.01\rightarrow2.00$  & clear & 1.11 & 2.49 &  \\
     &                       &        & CAO 1.2 m & $R_{\rm{C}}$ & $1.00\rightarrow2.08$  & clear & 1.33 & 1.85 &  \\
24 & 2011 Sep 08 & 904 & OSN 1.5 m & $R_{\rm{C}}$ & $1.01\rightarrow1.64$  & clear & 2.14 & 1.46 & $\surd$ \\
25 & 2011 Oct 02 & 917 & OSN 1.5 m & $R_{\rm{C}}$ & $1.05\rightarrow2.14$  & clear & 3.53 & 1.04 & $\surd$ \\
     &                       &       & Trebur 1.2 m & $R_{\rm{B}}$ & $1.13\rightarrow2.39$  & partly cloudy & 1.82 & 1.80 &  \\
     &                       &       & Jena 0.9 m & $R_{\rm{B}}$ & $1.13\rightarrow2.82$  & clear & 1.29 & 2.76 & \\
26 & 2011 Oct 15  & 924 & Herg.-Hall. 0.2 m & clear & $1.08\rightarrow1.88$  & clear & 0.62 & 3.44 & \\
\enddata
\tablecomments{Date UT is given for a mid-transit time, epoch $E$ is a transit number from the initial ephemeris given in \citet{Pollacco08}, $R_{\rm{C}}$ and $R_{\rm{B}}$ denote Cousins and Bessell $R$-band filters, respectively, $\Gamma$ is a median number of exposures per minute, $pnr$ is a photometric scatter in millimag per minute (see Sect.~\ref{Sect.TransitObs} for details). Light curves which were used to produce a transit template are marked with $\surd$ in the last column (Templ.).}
\end{deluxetable}

\clearpage

\begin{deluxetable}{ccc}
\tabletypesize{\scriptsize}
\tablecaption{New radial velocities for the WASP-3 system.\label{Tab.RVs}}
\tablewidth{0pt}
\tablehead{
\colhead{BJD$_{\rm{TDB}}$} & \colhead{RV (m~s$^{-1}$)} & \colhead{Error (m~s$^{-1}$)} 
}
\startdata
2455345.782701 &  195.802 &   36.312 \\
2455349.755560 &   -6.370 &   41.495 \\
2455350.765352 &  -71.253 &   30.760 \\
2455365.710214 &   92.082 &   36.094 \\
2455365.932853 &  238.030 &   40.097 \\
2455369.705307 &  196.649 &   37.140 \\
2455371.682112 &  186.664 &   28.075 \\
2455372.681765 & -262.104 &   38.049 \\
2455806.729146 & -243.796 &   31.687 \\
2455807.742803 &   -9.394 &   54.541 \\
2455810.717317 &   -1.821 &   33.684 \\
2455812.716356 &  117.557 &   37.807 \\
2455816.704736 &   93.935 &   32.336 \\
2455823.683879 &   76.051 &   33.085 \\
2455824.679898 & -301.463 &   33.483 \\
2455828.659585 & -409.340 &   36.002 \\
2455829.669701 &  180.807 &   41.666 \\
\enddata
\tablecomments{Times are given for a middle of exposure as BJD based on TDB. RV values are relative to the mean value representative for a whole dataset. All values are intentionally left unrounded.}
\end{deluxetable}

\clearpage

\begin{deluxetable}{lcccc}
\tabletypesize{\scriptsize}
\tablecaption{Parameters of the WASP-3 star derived from spectroscopic analysis.\label{Tab.SpecResults}}
\tablewidth{0pt}
\tablehead{
\colhead{Parameter} & \colhead{This work} & \colhead{Pol08}  & \colhead{Mon12}  & \colhead{Tor12}
}
\startdata
Effective temperature, $T_{\rm{eff}}$ (K)						& $6340\pm90$ 					& $6400\pm100$ & $6448\pm123$ & $6375\pm63$\\
Surface gravity, $\log g_{*}$ (cgs units)    					& $4.25\pm0.15$ 					& $4.25\pm0.05$ & $4.49\pm0.08$ & $4.28\pm0.03$\tablenotemark{a}\\
Metallicity, [Fe/H]						     				& $-0.161\pm0.063$ 				& $0.00\pm0.20$ & $-0.02\pm0.08$ & $-0.06\pm0.08$\\
Lithium abundance, $A(\rm{Li})$ (dex)						& $2.65\pm0.08$ & $2.0-2.5$ & &\\
Micro turbulent velocity, $v_{\rm{mic}}$  (km~s$^{-1}$)			& $1.4\pm0.3$ 					&  & &\\
Mass, $M_{*}$  ($M_{\odot}$)     							& $1.11^{+0.08}_{-0.06}$ 			& $1.24^{+0.06}_{-0.11}$ & &\\
Luminosity, $L_{*}$  ($L_{\odot}$)     						& $2.4^{+0.4}_{-0.3}$ 				& & &\\
Age, (Gyr)     										& $3.9^{+1.3}_{-1.2}$ 				& $0.7-3.5$ & &\\
Rotation velocity, $v \sin I$ (km~s$^{-1}$)		   			& $15.6\pm1.5$ 							& $13.4\pm1.5$ & $13.4-14.8$ & $15.4\pm1.2$\\
\enddata
\tablecomments{References: Pol08 - \citet{Pollacco08} , Mon12 - \citet{Montalto12}, Tor12 - \citet{Torres12}.}
\tablenotetext{a}{Based on photometric analysis}
\end{deluxetable}

\clearpage

\begin{deluxetable}{lcccccc}
\tabletypesize{\scriptsize}
\rotate
\tablecaption{Parameters of the WASP-3 system derived from transit light curve analysis.\label{Tab.TranResults}}
\tablewidth{0pt}
\tablehead{
\colhead{Parameter} & \colhead{This work} & \colhead{Pol08}  & \colhead{Gib08}  & \colhead{Chr11}  & \colhead{Mon12}  & \colhead{Nas13}
}
\startdata
Orbital inclination, $i_{\rm{b}}$ (degrees)               			& $84.15^{+0.41}_{-0.40}$  			& $84.4^{+2.1}_{-0.8}$ & $85.06^{+0.16}_{-0.15}$ & $84.22\pm0.81$ & & $84.12\pm0.82$\\
Scaled semimajor-axis, $a_{\rm{b}}/R_{*}$             			& $5.05^{+0.09}_{-0.09}$  			& & & & &\\
Planetary to stellar radii ratio, $R_{\rm{b}}/R_{*}$				& $0.10649^{+0.00061}_{-0.00063}$  & $0.1030^{+0.0010}_{-0.0015}$ & $0.1014^{+0.0010}_{-0.0008}$ & $0.1051\pm0.0124$ & $0.1061\pm0.0007$ & $0.1058\pm0.0012$ \\
Transit parameter, $b=\frac{a_{\rm{b}}}{R_{*}}\cos{i_{\rm{b}}}$ 	& $0.492^{+0.042}_{-0.041}$  		& $0.505^{+0.051}_{-0.166}$ & $0.448^{+0.014}_{-0.014}$ & &  & \\
Linear LD coefficient in $R$, $u_{\rm{R}}$             				& $0.24\pm0.04$  & & & & & $0.247^{+0.029}_{-0.028}$\\
Linear LD coefficient in $I$, $u_{\rm{I}}$                				& $0.23\pm0.10$  & & & & &\\
Linear LD coefficient in $r'$, $u_{r'}$              				& $0.28\pm0.06$  & & & & &\\
Linear LD coefficient in $i'$, $u_{i'}$              				& $0.18\pm0.06$  & & & & &\\
Planetary radius, $R_{\rm{b}}$ ($R_{\rm{Jup}}$)      				& $1.346 \pm 0.063$ 					& $1.31^{+0.07}_{-0.14}$ & $1.29^{+0.05}_{-0.12}$ & $1.385\pm0.060$ & &\\
Planetary density, $\rho_{\rm{b}}$ ($\rho_{\rm{Jup}}$)      		& $0.73^{+0.15}_{-0.14}$ 					& $0.78^{+0.28}_{-0.09}$ & $0.82^{+0.14}_{-0.09}$ & & &\\
Planetary surface gravity, $g_{\rm{b}}$ (m~s$^{-2}$)      			& $24 \pm 2$ 					& $23.4^{+5.4}_{-2.1}$ & $26.3^{+3.9}_{-2.3}$ & & &\\
Stellar radius, $R_{*}$  ($R_{\odot}$)                     				& $1.298^{+0.053}_{-0.045}$ 			& $1.31^{+0.05}_{-0.12}$ & $1.354\pm0.056$ & & &\\
Stellar density, $\rho_{*}$  ($\rho_{\odot}$)          				& $0.506^{+0.026}_{-0.025}$ 		& $0.55^{+0.15}_{-0.05}$ & & & $0.57\pm0.05$ &\\
Stellar surface gravity, $\log g_{*}$ (cgs units)     				& $4.26\pm0.04$ 			& $4.30^{+0.07}_{-0.03}$ & & & & \\
\enddata
\tablecomments{References: Pol08 - \citet{Pollacco08}, Gib08 - \citet{Gibson08}, Chr11 - \citet{Christiansen11}, Mon12 - \citet{Montalto12}, Nas13 - \citet{Nascimbeni13}.}
\end{deluxetable}

\clearpage

\begin{deluxetable}{lccll}
\tabletypesize{\scriptsize}
\tablecaption{Mid-transit times determined for individual epochs.\label{Tab.Timing}}
\tablewidth{0pt}
\tablehead{
\colhead{Date UT} & \colhead{$N_{\rm{LC}}$} & \colhead{Epoch} & \colhead{$T_{\rm{mid}}$ (BJD$_{\rm{TDB}}$)}  & \colhead{$O-C$ (d)}
}
\startdata
2009 Sep 24 & 1 & 517 & $2455098.66406\pm0.00044$ & $-0.00080$ \\
2010 May 25 & 1 & 649 & $2455342.4470\pm0.0012$ & $-0.0001$ \\
2010 Jun 07 & 2 & 656 & $2455355.37419\pm0.00053$ & $-0.00075$ \\
2010 Jun 15 & 2 & 660 & $2455362.76233\pm0.00040$ & $+0.00005$ \\
2010 Jun 18 & 1 & 662 & $2455366.4561\pm0.0010$ & $+0.0002$ \\
2010 Jun 25 & 1 & 666 & $2455373.84230\pm0.00086$ & $-0.00099$ \\
2010 Jul 09 & 1 & 673 & $2455386.770\pm0.0020$ & $-0.0012$ \\
2010 Jul 24 & 1 & 681 & $2455401.54564\pm0.00036$ & $-0.00017$ \\
2010 Aug 05 & 1 & 688 & $2455414.47368\pm0.00073$ & $+0.00002$ \\
2010 Aug 28 & 1 & 700 & $2455436.6353\pm0.0010$ & $-0.0004$ \\
2010 Aug 29 & 1 & 701 & $2455438.4827\pm0.0006$ & $+0.0002$ \\
2010 Sep 11 & 2 & 708 & $2455451.4101\pm0.0004$ & $-0.0003$ \\
2011 Apr 03 & 1 & 818 & $2455654.5618\pm0.0014$ & $-0.0005$ \\
2011 May 09 & 1 & 838 & $2455691.49938\pm0.00086$ & $+0.00045$ \\
2011 May 17 & 1 & 842 & $2455698.88641\pm0.00027$ & $+0.00015$ \\
2011 Jun 03 & 1 & 851 & $2455715.50824\pm0.00072$ & $+0.00046$ \\
2011 Jun 06 & 1 & 869 & $2455748.7507\pm0.0011$ & $-0.0001$ \\
2011 Jun 15 & 1 & 858 & $2455728.43608\pm0.00052$ & $+0.00045$ \\
2011 Jun 27 & 1 & 864 & $2455739.51735\pm0.00064$ & $+0.00071$ \\
2011 Jul 22 & 1 & 878 & $2455765.3718\pm0.0014$ & $-0.0005$ \\
2011 Aug 02 & 1 & 884 & $2455776.45452\pm0.00092$ & $+0.00118$ \\
2011 Aug 14 & 1 & 890 & $2455787.53583\pm0.00053$ & $+0.00147$ \\
2011 Aug 26 & 2 & 897 & $2455800.46137\pm0.00055$ & $-0.00083$ \\
2011 Sep 08 & 1 & 904 & $2455813.3891\pm0.0015$ & $-0.0009$ \\
2011 Oct 02 & 3 & 917 & $2455837.39876\pm0.00044$ & $-0.00014$ \\
2011 Oct 15 & 1 & 924 & $2455850.3274\pm0.0010$ & $+0.0006$ \\
\enddata
\tablecomments{Date UT is given for a mid-transit time, $N_{\rm{LC}}$ is a number of light curves used, epoch is a transit number from the initial ephemeris, $T_{\rm{mid}}$ is mid-transit time in BJD based on TDB, and $O-C$ is is the timing deviation from the linear ephemeris.}
\end{deluxetable}

\clearpage

\begin{deluxetable}{lccc}
\tabletypesize{\scriptsize}
\tablecaption{Orbital parameters for the WASP-3 b planet.\label{Tab.OrbiResults}}
\tablewidth{0pt}
\tablehead{
\colhead{Parameter} & \colhead{This work} & \colhead{Pol08}  &  \colhead{Mon12}
}
\startdata
RV semi-amplitude, $k$ (m~s$^{-1}$)               				& $272\pm10$ 					& $251.2^{+7.9}_{-10.8}$ & $277-290$  \\
Semi-major axis, $a_{\rm{b}}$ (AU)               					& $0.0305^{+0.0008}_{-0.0006}$ & $0.0317^{+0.0005}_{-0.0010}$ & \\
Minimum planetary mass, $M_{\rm{b}} \sin i_{\rm{b}}$ ($M_{\rm{Jup}}$) 	& $1.76^{+0.11}_{-0.09}$ 				& & \\
Planetary mass, $M_{\rm{b}}$ ($M_{\rm{Jup}}$)      				& $1.77^{+0.11}_{-0.09}$ 				& $1.76^{+0.08}_{-0.14}$ & \\
\enddata
\tablecomments{References: Pol08 - \citet{Pollacco08}, Mon12 - \citet{Montalto12}.}
\end{deluxetable}


\begin{thebibliography}{}
  \bibitem[Adam\'ow et al. (2012)]{Adamow12} Adam\'ow, M., Niedzielski, A., \& Wolszczan, A. 2012, Mem. S.A.It. Suppl., 22, 48
  \bibitem[Borucki et al. (2010)]{Borucki} Borucki, W. J., Koch, D. G., Basri, G., et al. 2010, Sci, 327, 977  
  \bibitem[Bressan et al. (2012)]{Bressan12} Bressan, A., Marigo, P., Girardi, L., et al. 2012, \mnras, 427, 127
  \bibitem[Carter \& Winn (2009)]{Carter09} Carter, J.A., \& Winn, J.N. 2009, \apj, 704, 51
  \bibitem[Carter et al. (2011)]{Carter11} Carter, J.A., Winn, J.N., Holman, et al. 2011, \apj, 730, 82
  \bibitem[Chambers (1999)]{Chambers99} Chambers, J. E. 1999, \mnras, 304, 793
  \bibitem[Christiansen et al. (2011)]{Christiansen11} Christiansen, J. L., Ballard, S., Charbonneau, D., et al. 2011, \apj, 726, 94
  \bibitem[Claret \& Bloemen (2011)]{Claret11} Claret, A., \& Bloemen, S. 2011, \aap, 529, A75
  \bibitem[Csizmadia et al. (2012)]{Csizmadia12} Csizmadia, Sz., Pasternacki, Th., Dreyer, C., et al. 2012, \aap, 549, A9
  \bibitem[Cutri et al. (2003)]{Cutri03} Cutri, R. M., Skrutskie, M. F., van Dyk, S., et al. 2003, VizieR Online Data Catalog, 2246, 0
  \bibitem[Eastman et al.  (2010)]{Eastman} Eastman, J., Siverd, R., \& Gaudi, B. S. 2010, \pasp, 122, 935
  \bibitem[Eastman et al. (2013)]{Eastman12} Eastman, J., Gaudi, B. S., \& Agol, E.  2013, \pasp, 125, 83
  \bibitem[Eibe et al. (2012)]{Eibe12} Eibe, M. T., Cuesta, L., Ull\'an, A., P\'erez-Verde, A., \& Navas, J. 2012, \mnras, 423, 1381
  \bibitem[Fulton et al. (2011)]{Fulton} Fulton, B. J., Shporer, A., Winn, J. N., et al. 2011, \aj, 142, 84
  \bibitem[Gazak et al. (2012)]{Gazak12} Gazak, J. Z., Johnson, J. A., Tonry, J., et al. 2012, Advances in Astronomy, 2012, 697967
  \bibitem[Gibson et al. (2008)]{Gibson08} Gibson, N. P., Pollacco, D., Simpson, E. K., et al. 2008, \aap, 492, 603  
  \bibitem[Gomes da Silva et~al. (2012)]{2012AA...541A...9G} Gomes da Silva, J., Santos, N.~C., Bonfils, X., et al. 2012, \aap, 541, A9
  \bibitem[H{\o}g \& Murdin (2000)]{Hog00} H{\o}g, E., \& Murdin, P. 2000, Tycho Star Catalogs: The 2.5 Million Brightest Stars, ed. Murdin, P.
  \bibitem[Howarth (2011)]{Howarth11} Howarth, I. D. 2011, \mnras, 418, 1165
  \bibitem[Hoyer et al. (2012)]{Hoyer12} Hoyer, S., Rojo, P., \& L\'opez-Morales, M. 2012, \apj, 748, 22
  \bibitem[Latham et al. (2011)]{Latham11} Latham, D. W., Rowe, J. F., Quinn, S. N., et al. 2011, \apj, 732, L24
  \bibitem[Littlefield (2011)]{Littlefield11} Littlefield, C. 2011, arXiv:1106.4312
  \bibitem[Lomb (1976)]{Lomb} Lomb, N. R. 1976, Ap\&SS, 39, 447
  \bibitem[Maciejewski et al. (2010)]{Maciejewski10} Maciejewski, G., Dimitrov, D., Neuh\"auser, R., et al. 2010, \mnras, 407, 2625
  \bibitem[Maciejewski et al. (2013)]{Maciejewski13} Maciejewski, G., Dimitrov, D., Seeliger, M., et al. 2013, \aap, 551, A108
  \bibitem[Mandel \& Agol (2002)]{MandelAgol} Mandel, K., \& Agol, E. 2002, \apjl, 580, L171
  \bibitem[McLaughlin (1924)]{McLaughlin} McLaughlin, D. B. 1924, \apj, 60, 22
  \bibitem[Meschiari et al. (2009)]{Meschiari09} Meschiari, S., Wolf, A. S., Rivera, E., et al. 2009, \pasp, 121, 1016
  \bibitem[Miller et al. (2010)]{Miller10} Miller, G. R. M., Collier Cameron, A., Simpson, E. K., et al. 2010, \aap, 523, A52
  \bibitem[Molaro \& Monai (2012)]{2012AA...544A.125M} Molaro, P., \& Monai, S. 2012, \aap, 544, A125
  \bibitem[Montalto et al. (2012)]{Montalto12} Montalto, M., Gregorio, J., Bou\'e, A., et al. 2012, \mnras, 427, 2757
  \bibitem[Mugrauer \& Berthold (2010)]{mugrauer} Mugrauer, M., \& Berthold, T. 2010, Astron. Nachr., 331, 449
  \bibitem[Nascimbeni et al. (2013)]{Nascimbeni13} Nascimbeni, V., Cunial, A., Murabito, S., et al. 2013, \aap, 549, A30
  \bibitem[Neuh\"auser et al. (2011)]{Neuhauser11} Neuh\"auser, R., Errmann, R., Berndt, A., et al. 2011, Astron. Nachr., 332, 547
\bibitem[Niedzielski et~al. (2011)]{2011IAUS..276..445N} Niedzielski, A., Wolszczan, A., Nowak, G., Zieli{\'n}ski, P., Adam{\'o}w, M., \& Gettel, S. 2011, in A. {Sozzetti}, M.~G. {Lattanzi}, and A.~P. {Boss} (eds.), {\em IAU Symposium}, Vol. 276 of {\em IAU Symposium}, pp 445--447
 \bibitem[Nowak et al. (2013)]{Nowaketal2013} {Nowak}, G., {Wolszczan}, A., {Niedzielski}, A., {Adam{\'o}w}, M., \& {Maciejewski}, G. 2013, \apj, 770, 53
  \bibitem[Pollacco et al. (2008)]{Pollacco08} Pollacco, D., Skillen, I., Collier Cameron, A., et al. 2008, \mnras, 385, 1576
  \bibitem[Ram\'irez \& Mel\'endez (2005)]{Ramirez05} Ram\'irez, I., \& Mel\'endez, J. 2005, \apj, 626,
465
  \bibitem[Ramsey et al. (1998)]{Ramsey98} Ramsey, L. W., Adams, M. T., Barnes, T. G., et al. 1998, in Society of Photo-Optical Instrumentation Engineers (SPIE) Conference Series, ed. L. M. Stepp, Vol. 3352, 34–42
  \bibitem[Robertson  et~al. (2013)]{2013ApJ...764....3R} Robertson, P., Endl, M., Cochran, W.~D., \& Dodson-Robinson, S.~E.  2013, \apj, 764, 3
  \bibitem[Rossiter (1924)]{Rossiter} Rossiter, R. A. 1924, \apj, 60, 15
  \bibitem[Sada et al. (2012)]{Sada12} Sada, P. V., Deming, D., Jennings, D. E., et al. 2012, \pasp, 124, 212
  \bibitem[Scargle (1982)]{Scargle} Scargle, J. D. 1982, \apj, 263, 835
  \bibitem[Schneider (2000)]{Schneider00} Schneider, J. 2000, in ASP Conference Series, Vol. 212, From Giant Planets to Cool Stars, eds. C.A.Griffith \& M.S.Marley (San Francisco: ASP), p.284
  \bibitem[Sestito \& Randich (2005)]{Sestito05} Sestito, P., \& Randich, S. 2005, \aap, 442, 615
  \bibitem[Silva (2003)]{Silva03} Silva, A. 2003, \apj, 585, L147
  \bibitem[Simpson et al. (2010)]{Simpson10} Simpson, E. K., Pollacco, D., H\'ebrard, G., et al. 2010, \mnras, 405, 1867
  \bibitem[Sousa et al. (2007)]{Sousa07} Sousa, S.~G., Santos, N.~C., Israelian, G., Mayor, M., \& Monteiro, M.~J.~P.~F.~G. 2007, \aap, 469, 783
  \bibitem[Southworth et al. (2004a)]{Southwortha} Southworth, J., Maxted, P. F. L., \& Smalley, B. 2004a, \mnras, 349, 547
  \bibitem[Southworth (2004b)]{Southworthb} Southworth, J., Maxted, P. F. L., \& Smalley, B. 2004b, \mnras, 351, 1277
  \bibitem[Steffen et al. (2012)]{Steffen12} Steffen, J. H., Ragozzine, D., Fabrycky, D. C., 2012, PNAS, 109, 7982
  \bibitem[Straizys \& Kuriliene (1981)]{Straizys81} Straizys, V., \& Kuriliene, G. 1981, Ap\&SS, 80, 353
  \bibitem[Stumpff (1980)]{1980AAS...41....1S} Stumpff, P. 1980, \aaps, 41, 1
  \bibitem[Takeda  et al. (2002)]{Takeda02} Takeda, Y., Ohkubo, M., \& Sadakane, K. 2002, \pasj, 54, 451
  \bibitem[Takeda  et al. (2005)]{Takeda05} Takeda, Y., Ohkubo, M., Sato, B., Kambe, E., \& Sadakane, K. 2005, \pasj, 57, 27
  \bibitem[Torres  et al. (2012)]{Torres12} Torres, G., Fischer, D. A., Sozzetti, A., et al. 2012, \apj, 757, 161
  \bibitem[Tripathi et al. (2010)]{Tripathi10} Tripathi, A., Winn, J. N., Johnson, J. A., et al. 2010, \apj, 715, 421
  \bibitem[Tull (1998)]{Tull98} Tull, R. G. 1998, in Society of Photo-Optical Instrumentation Engineers (SPIE) Conference Series, ed. S. D'Odorico, Vol. 3355, 387
  \bibitem[Valenti \& Piskunov (1996)]{Valenti96} Valenti, J. A., \& Piskunov, N. 1996, A\&AS, 118, 595
  \bibitem[Zhao et al. (2012)]{Zhao12} Zhao, M., Milburn, J., Barman, T., et al. 2012, \apjl, 748, 8
\end{thebibliography}
\end{document}